\documentclass[fleqn,usenatbib]{mnras}

\usepackage{newtxtext,newtxmath}

\usepackage[T1]{fontenc}

\usepackage{graphicx}	
\usepackage{amsmath}	
\usepackage{tabularx}

\title[Constraints on the companion of CQ Tau]{External or internal companion exciting the spiral arms in CQ Tau?}

\author[I. Hammond et al.]{Iain Hammond$^{1}$\thanks{E-mail: \href{mailto:Iain.Hammond@monash.edu}{Iain.Hammond@monash.edu}}, 
Valentin Christiaens$^{1,2}$, 
Daniel J. Price$^{1}$, 
Maria Giulia Ubeira-Gabellini$^{3}$,
\newauthor{Jennifer Baird$^{1}$,
Josh Calcino$^{4}$,
Myriam Benisty$^{5,6}$}, 
Giuseppe Lodato$^{3}$,
Leonardo Testi$^{7,8}$,
\newauthor{Christophe Pinte$^{1,5}$, 
Claudia Toci$^{3}$ and
Davide Fedele$^{8,9}$}
\\
$^{1}$ School of Physics and Astronomy, Monash University, Vic 3800, Australia\\
$^{2}$ Space sciences, Technologies \& Astrophysics Research (STAR) Institute, Universit\'e de Li\`ege, All\'ee du Six Ao\^ut 19c, B-4000 Sart Tilman, Belgium\\
$^{3}$ Dipartimento di Fisica, Universit{\`a} Degli Studi di Milano, Via Celoria, 16, Milano, I-20133, Italy\\
$^{4}$ Theoretical Division, Los Alamos National Laboratory, Los Alamos, NM 87545, USA\\
$^{5}$ Univ. Grenoble Alpes, CNRS, IPAG, F-38000 Grenoble, France\\
$^{6}$ Unidad Mixta Internacional Franco-Chilena de Astronomía (CNRS, UMI 3386), Departamento de Astronomía, \\Universidad de Chile, Camino El Observatorio 1515, Las Condes, Santiago, Chile\\
$^{7}$ European Southern Observatory (ESO), Karl-Schwarzschild-Str 2, D-85748 Garching, Germany\\
$^{8}$ INAF, Osservatorio Astrofisico di Arcetri, Largo Enrico Fermi 5, 50125, Firenze, Italy\\
$^{9}$ INAF, Osservatorio Astrofisico di Torino, Via Osservatorio 20, 10025, Pino Torinese, Italy\\
}

\date{Accepted 2022 July 25. Received 2022 July 19; in original form 2022 February 21}

\pubyear{2022}

\begin{document}
\label{firstpage}
\pagerange{\pageref{firstpage}--\pageref{lastpage}}
\maketitle

\begin{abstract}

We present new high-contrast images in near-infrared wavelengths ($\lambdaup$\textsubscript{$c$}~=~1.04, 1.24, 1.62, 2.18 and 3.78$\mu$m) of the young variable star CQ~Tau, aiming to constrain the presence of companions in the protoplanetary disc. We reached a \textit{Ks}-band contrast of 14 magnitudes with SPHERE/IRDIS at separations greater than 0\farcs4 from the star. Our mass sensitivity curve rules out giant planets above 4~M$_{\rm Jup}$ immediately outside the spiral arms at $\sim$60 au and above 2--3~M$_{\rm Jup}$ beyond 100 au to 5$\sigma$ confidence assuming hot-start models. We do, however, detect four spiral arms, a double-arc and evidence for shadows in scattered light cast by a misaligned inner disc. Our observations may be explained by an unseen close-in companion on an inclined and eccentric orbit. Such a hypothesis would also account for the disc CO cavity and disturbed kinematics. 

\end{abstract}




\section{Introduction} \label{Intro}

The advent of dedicated high-contrast facilities, such as the near-infrared (NIR) Spectro-Polarimetic High contrast imager for Exoplanets REsearch \citep[SPHERE;][]{Beuzit2008} instrument and the Atacama Large Millimeter Array (ALMA), triggered a revolution in imaging of protoplanetary discs. The enhanced angular resolution and sensitivity has led to the discovery of substructures in protoplanetary discs, such as annular gaps \citep[e.g.,][]{Andrews2018, Long:2018tl}, inner cavities \citep[e.g.,][]{Canovas2015,Villenave:2019uy}, crescent-shape asymmetries \citep[e.g.,][]{Casassus2013}, spiral arms \citep[e.g.,][]{Muto2012, Rosotti:2020ul} and shadows \cite[e.g.,][]{Keppler:2020vi}, interpreted as possible signposts for the presence of embedded companions.  

Hydrodynamical simulations of perturbing bodies in protoplanetary discs have had some success in recreating observed substructures. A few examples of discs where these models have been applied are HD~142527 \citep[e.g.,][]{Avenhaus:2014vu,Price:2018tu}, MWC~758 \citep[e.g.,][]{Benisty2015,Dong2015a,Reggiani2018,Calcino2020}, IRS~48 \citep[e.g.,][]{van-der-Marel2013,vanderMarel2016b,Calcino2019}, HD~100453 \citep[e.g.,][]{Benisty2017,Rosotti:2020ul,Gonzalez2020}, HD~169142 \citep[e.g.,][]{Bertrang2018,Toci:2020aa}, HD~100546 \citep[e.g.,][]{Follette2017,Fedele:2021uu} and PDS~70
\citep[e.g.,][]{Keppler2018,Toci:2020wu}. In some cases, different companions have been proposed to explain the same substructures (e.g., for MWC~758 from \citealt{Calcino2020} and \citealt{Baruteau2019}).

Spiral arms have long been hypothesised to arise due to planet-disc interactions \citep{Goldreich1979, Goldreich1980, Ogilvie2002}. The amplitude of the spirals depends on the mass of the planet with respect to the so-called thermal mass (the mass for which the Hill radius equals the local gas scale height) 
and properties of the disc such as the temperature profile. In general, low mass planets (e.g., masses much lower than the thermal mass
) are expected to produce a single inner and outer wake. When the planetary mass increases to approximately the thermal mass or higher, multiple spirals are generated \citep{Bae:2018wb} both interior and exterior to the planet. Perturbations from such a massive planet external to the spiral arms initially seemed to be an exciting and robust explanation for the apparent $m=2$ spirals detected in several protoplanetary discs \citep[most notably MWC~758 and HD~135344B,][]{Garufi2013, Benisty2015}. However, no convincing detection has emerged, despite mass limits of the order of 3~M$_{\rm Jup}$ exterior to the spirals assuming hot-start formation scenario for the cases of HD~135344B and MWC~758 \citep{Maire2017,Reggiani2018,Boccaletti:2021tl}.

Radial velocity and direct imaging surveys have found super-Jupiter mass planets with non-negligible eccentricities \citep{Bitsch2020}. If these eccentric orbits were present during the protoplanetary disc phase the resulting spiral structure would be drastically different to the spiral morphology expected from perturbers on circular orbits \citep{Zhu:2022vw}. \cite{Calcino2020} showed that a super-Jupiter mass planet on an eccentric orbit can produce multiple spiral arms outside of a central cavity. Several studies \citep[e.g.,][]{Muley2019, Baruteau2021} have found massive planets become eccentric as they begin to carve deep gaps which deplete the co-orbital Lindblad torques that tend to quickly damp any growth in eccentricity \citep{Goldreich1980, Artymowicz1993}. In this context, the peculiar spiral morphology of CQ~Tau \citep{Uyama2020} makes it a promising laboratory to better understand the origin of spiral arms in protoplanetary discs.

CQ~Tau is a young ($\sim$9 Myr, \citealt{Vioque2018}) variable star of intermediate mass (1.47~M$_{\odot}$, \citealt{Vioque2018}), located near the Taurus star forming region (RA = 05h:35m:58.47s, Dec = +24$^{\circ}$44:54.09; J2000) at a distance of 149~pc \citep{Gaia-Collaboration:2021tu}. It is spectral type F2 (see Table \ref{tab:CQ_Tau_prop}) and shows UX Ori-type irregular photometric variability, possibly indicative of the presence of a companion (Section \ref{sec:binarity}). 
CQ~Tau is surrounded by a massive circumstellar disc, historically observed in 
continuum emission with multiple instruments (OVRO interferometer, \citealt{Mannings1997}; PdBI, \citealt{Natta2000}; VLA, \citealt{Testi2001b}). Its Spectral Energy Distribution (SED) shows infrared and mm-wavelength emission, allowing us to infer the presence of grains larger than those in the interstellar medium (\citealt{Testi2001b,Testi2003, Chapillon2008}), up to cm-sized (\citealt{Banzatti2011}). There is evidence of grain growth variation with larger grains present in the inner disc \citep{Trotta2013}. Moreover, CQ~Tau has a relatively high accretion rate of the order of 10$^{-8}$--10$^{-7}$M$_\odot$yr$^{-1}$ \citep{Donehew2011,Mendigutia2012, Guzman-Diaz:2021wj}. 

High resolution observations tracing the mm-dust (e.g., \citealt{Tripathi2017,Pinilla2018,Ubeira_Gabellini2019}) and gas distribution (e.g., \citealt{Ubeira_Gabellini2019,Wolfer2020,van-der-Marel:2021ww}) shows the presence of an inner cavity. \citet{Ubeira_Gabellini2019} analysed the dust and gas surface density profiles, finding a Gaussian dust ring at a radius of 53 au in the 1.3mm continuum (also \citealt{Francis:2020ve}) and a less prominent gas cavity of 20--25 au in the $^{13}$CO and C$^{18}$O emission. Based on hydrodynamical simulations, the authors suggested it to be caused by an undetected planet of mass 6--9~M$_{\rm Jup}$ located at $\sim$20 au. Recently, \citet{Uyama2020} presented \textit{L}$^{\prime}$-band Keck/NIRC2 and \textit{H}-band polarimetric Subaru/AO188+HiCIAO observations of CQ~Tau. In the small grain distribution they detected a spiral arm with a high pitch angle (34$\pm$2\degr), possibly induced by an unseen companion candidate, as well as a potential secondary spiral. 
No companion was detected, despite mass sensitivity limits of $\sim$7~M$_{\rm Jup}$ immediately beyond the prominent NIR spiral ($\sim$60 au) (\citealt{Uyama2020}). This is in tension with hydrodynamical models \citep{Sanchis:2020wc} which predict a 1--2~M$_{\rm Jup}$ planet should be readily detectable at radii $\gtrsim$5--10 au with a contrast of 14--16 magnitudes respectively, due to negligible disc extinction in \textit{L}$^{\prime}$-band. Finally, \citet{Wolfer2020} analysed the gas brightness temperature and kinematics of the disc, finding non-Keplerian gas substructures, such as three spiral structures spanning radii from $\sim$10--180 au and connected with the spirals observed in NIR scattered light, and a nearly symmetric temperature dip approximately co-spatial with the semi-minor axis of the disc. 

In this paper, we first present new NIR images and constraints on possible companions in the outer disc of CQ~Tau. We present our data reduction and post-processing in Section~\ref{Obs+DataRed}, our spiral arm characterisation and detection limits in Section~\ref{sec:results}, and in Section~\ref{Discussion} we discuss the indicators of a companion in the disc of CQ~Tau.


\newcommand\T{\rule{0pt}{2.6ex}}       
\newcommand\B{\rule[-1.2ex]{0pt}{0pt}} 
\begin{table}
\caption{Stellar and disc properties of CQ Tau.}  
\label{tab:CQ_Tau_prop}
\begin{tabularx}{0.45\textwidth}{    
   >{\raggedright\arraybackslash}X   
   >{\centering\arraybackslash}X     
   >{\centering\arraybackslash}X}    
\hline 
\hline
Parameter\T\B            & Value\T\B                & Reference\T\B\\ 
\hline
Distance [pc]            & $149.37^{+1.35}_{-1.33}$ & 1\T\B\\
Age [Myrs]               & $8.9^{+2.8}_{-2.5}$      & 2\\
Spectral Type            & F2                       & 3, 4\\
T\textsubscript{eff} [K] & 6850$\pm$100             & 2, 5\\
Av [mag]                 & 0.41                     & 2\\
Log(L) [L$_\odot$]       & 0.87                     & 2\\
M [~M$_\odot$]            & 1.47                     & 2\\
\hline
\end{tabularx}
\vspace{0.2cm}

References: (1) \citet[][]{Gaia-Collaboration:2021tu}, (2) \citet[][]{Vioque2018}, \\ (3) 
\citet[][]{Herbig1960}, (4)
\citet{Natta2001}, (5) 
\citet{Dodin:2021tj}.
\end{table}

\section{Observations and data reduction}\label{Obs+DataRed}
\begin{table*} 
\begin{center}
\caption{Summary of the archival VLT high-contrast imaging observations on CQ Tau used in this work.} 
\label{tab:CQTau_obs}
\begin{tabular}{lcccccccccccc}
\hline
\hline
Date & Strategy & Program & Instrument & Filter & Plate scale$^{\rm (a)}$ & Coronagraph  & 
T$_{\rm int}^{\rm (b)}$ & $<\beta>^{\rm (c)}$ & $\Delta$PA$^{\rm (d)}$ \\
& &  & &  & [mas px$^{-1}$] & & 
[min] & [\arcsec] & \degr \\ 
\hline
2016-12-21 & ADI & 298.C-5014 & IRDIS & $Ks$ & 12.265$\pm 0.009$  & N\_ALC\_YJH\_S  & 
127 & 0.69 & 44.4\\
2016-12-21 & A/SDI & 298.C-5014 & IFS & $YJH$ & 7.46$\pm 0.02$ & N\_ALC\_YJH\_S & 128 & 0.69 & 44.4 \\
2017-10-06 & PDI & 098.C-0760 & IRDIS & $J$ & 12.263$\pm 0.009$ & N\_ALC\_YJH\_S  & 
33 & 0.41 & -- \\
2018-02-18 & PDI & 098.C-0760 & IRDIS & $J$ & 12.238$\pm 0.009$ & -- & 20 & 0.65 & -- \\
2018-11-28 & ADI & 1101.C-0092 & NACO & $L^{\prime}$ & 27.208$\pm 0.088$ & AGPM & 157 & 0.55 & 67.4 \\
\hline
\end{tabular}
\end{center}
Notes: $^{\rm (a)}$\citet{Maire:2016vq} for IRDIS and IFS, \citet{launhardt2020} for NaCo. $^{\rm (b)}$Total integration time calculated after bad frame removal. $^{\rm (c)}$Average seeing at $\lambdaup$~=~500nm achieved during the sequence, returned by ESO's DIMM station. $^{\rm (d)}$Cumulative parallatic angle variation. 
\end{table*}

The observations used in this work are summarised in Table~\ref{tab:CQTau_obs}. All are new to this work, with the exception of program 098.C-0760 (2017-10-06) which is presented in Appendix E of \citet{Bohn:2022tx}. Together, they represent all currently available ESO high-contrast imaging data for CQ~Tau. Section \ref{sec:pdi} presents our PDI reduction, Section \ref{sec:irdis} our IRDIFS reduction and Section \ref{sec:naco} our NaCo reduction and new pipeline.

\subsection{VLT/SPHERE polarimetric data sets} \label{sec:pdi}
 Two polarimetric data sets were captured with the Infrared Differential Imaging Spectrometer \citep[IRDIS;][]{Dohlen:2008uh} instrument in dual-polarization mode \citep{de-Boer:2020vx, van-Holstein:2020vk} on October 6th, 2017 and February 18th, 2018 with the SPHERE instrument at the Very Large Telescope (VLT). 
 The exposure time was set to 50s for individual coronagraphic (apodised Lyot coronagraph \texttt{N\_ALC\_YJH\_S}, diameter $\sim$185 mas) images and 0.8375s for non-coronagraphic images providing a total of ten and 18 polarimetric cycles, respectively. The \textit{J}-band filter ($\lambdaup$\textsubscript{$c$}~=~1.24$\mu$m) was used for both observations.

Both data sets were reduced using the IRDIS Data reduction for Accurate Polarimetry pipeline \citep[\textsc{irdap} v1.3.3;][]{van-Holstein:2020tm, van-Holstein:2020vk}. This automated pipeline incorporates image calibration (dark subtraction, flat fielding and correction of bad pixels), pre-processing and post-processing for polarimetric data. The position of the central star was interpolated using the two center calibration files; one taken at the beginning of the observation and one at the end. For each polarimetric cycle, four cubes each corresponding to linear polarisation directions of \textit{Q\textsuperscript{+}, Q\textsuperscript{--}, U\textsuperscript{+}} and \textit{U\textsuperscript{--}} were obtained. Using these frames, the clean Stokes \textit{Q} and \textit{U} were determined using the standard double-difference method, corrected for instrument polarisation with the Mueller matrix model and converted to polar coordinates ($Q_{\phi}$ and $U_{\phi}$). By fitting a 2D Gaussian to the unsaturated flux frames of the coronagraphic data set, we measured a Full Width at Half Maximum (FWHM) of 
3.61 pixels after excluding one bad unsaturated cube.

We explored polarised phase function effects by producing two estimated total intensity images with the coronagraphic $Q_\phi$ image. The first using an in-house processing routine assuming typical grain sizes smaller than the wavelength of the observation (1.24$\mu$m). For independent confirmation, we used the \texttt{total\_intensity} and \texttt{phase\_function} routines in \textsc{diskmap} \citep{Stolker:2016ub} using our $r^{2}$-scaled image mapped with the scattering surface described in \citet{Ubeira_Gabellini2019}.

\subsection{VLT/SPHERE IRDIFS data set} \label{sec:irdis}

CQ~Tau was observed by SPHERE in \texttt{IRDIFS-EXT} mode on December 21st, 2016 (Program ID 298.C-5014(A), PI: L. Testi). This mode allowed the simultaneous acquisition of Integral Field Spectroscopy \citep[IFS;][]{Claudi:2008ut} and IRDIS data, in the $YJH$ ($\lambdaup$\textsubscript{$c$}~=~1.04, 1.24, 1.62$\mu$m) and $Ks$ ($\lambdaup$\textsubscript{$c$}~=~2.18$\mu$m) bands, respectively. Observations were made with a stabilised pupil to enable the application of angular differential imaging-based \citep[ADI,][]{Marois2006} post-processing algorithms to model and subtract the emission from the star. For both data sets the \texttt{N\_ALC\_YJH\_S} coronagraph (inner working angle $\sim$99 mas) was used and the total integration time was 128 minutes (Table \ref{tab:CQTau_obs}). IRDIS images had an individual exposure time of 16s, while IFS images were 32s each. For both IFS and IRDIS, images containing satellite spots \citep{Marois:2006tx,Langlois:2013uv} (labelled \texttt{CENTER}) were acquired every $\sim$31 minutes for a total of five \texttt{CENTER} cubes. 

\subsubsection{IFS}
The IFS data are acquired with a spectral resolution of R$\sim$30 and span a field of view of 1\farcs73 x 1\farcs73. We performed basic data calibration using the ESO SPHERE pipeline\footnote{\url{http://www.eso.org/sci/software/pipelines/sphere/}} (v0.36), involving dark subtraction, detector flat-fielding, anamorphism correction \citep{Maire:2016vq}, wavelength calibration, and spectral cube construction. 

Bad pixel correction and satellite spot-based centering of the images utilised routines implemented in the Vortex Image Processing\footnote{\url{https://github.com/vortex-exoplanet/VIP}} package \citep[\textsc{vip};][]{GomezGonzalez2017}. In order to increase the signal-to-noise (S/N) of the four satellite spots used for centering, we applied a high-pass filter to our \texttt{CENTER} images where the image itself was subtracted with a median low-pass filtered version of the image. We used a 2D Moffat function to fit the satellite spot positions and to derive the centroid of the star for each \texttt{CENTER} image. 
For the \texttt{OBJECT} images, the stellar position behind the coronagraph was interpolated using the \texttt{CENTER} images interspersed throughout the sequence.

We post-processed our calibrated, centered cubes using Principal Component Analysis functions implemented in \textsc{vip} (\citealt{amara2012,GomezGonzalez2017}). We used two flavours of PCA, leveraging either the spectral diversity alone (SDI) or both the spectral and angular diversity present in the data \citep[ASDI,][]{Ubeira_Gabellini_2020}.

\subsubsection{IRDIS}
IRDIS data consisted of 160 science cubes. We used an in-house reduction pipeline \citep{2021MNRAS.502.6117C} to reduce the data set, after adaptation to both support satellite spot centering and handle classical imaging data. The pipeline uses ESO's Common Pipeline library \textsc{esorex} (v3.13.5) recipes for calibration, and \textsc{vip} functions for pre- and post-processing. The breakdown of the pipeline for this observation is as follows: i) dark current subtraction, ii) flat-fielding iii) PCA-based sky subtraction, iv) bad pixel correction, v) centering via satellite spots, vi) anamorphism correction, vii) finer centering using a background star (see Appendix \ref{sec:astrometry}), viii) bad frame rejection (two frames), ix) final PSF creation, x) cropping, xi) post-processing with median-ADI, PCA-ADI in full frame and annular PCA-ADI. 
We measured a FWHM of 5.24 pixels.

\begin{figure*}
	\centering
	\includegraphics[width=0.92\textwidth]{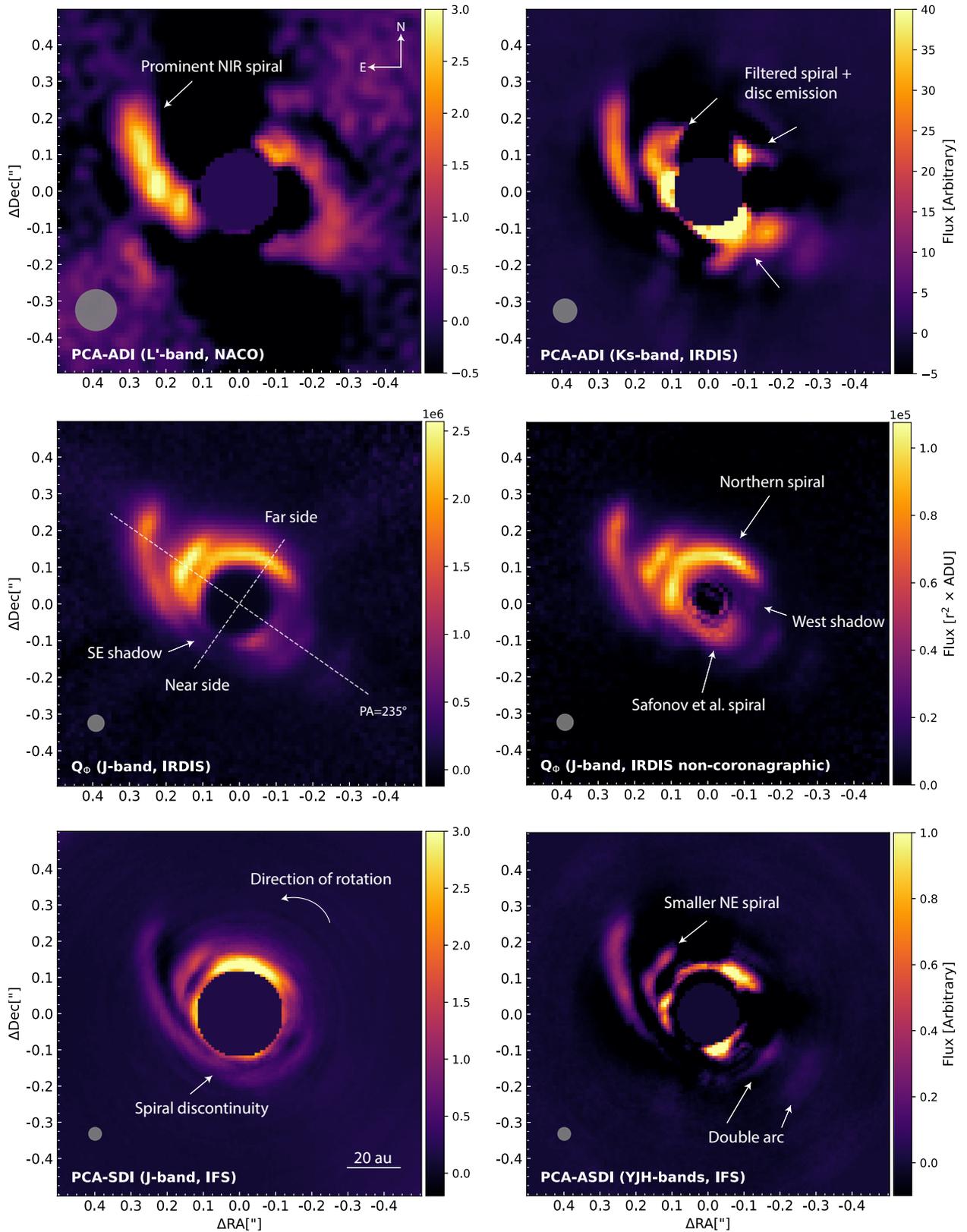}

	\caption{Our reduced and annotated multi-wavelength near-IR scattered light observations of CQ~Tau. 
	From left to right, top to bottom: VLT/NaCo ADI (\textit{L}$^{\prime}$-band, npc=1), SPHERE/IRDIS ADI (\textit{Ks}-band, npc=1), SPHERE/IRDIS PDI (\textit{J}-band, $r^2$-scaling), SPHERE/IFS SDI (\textit{J}-band, npc=1) and ASDI (\textit{YJH}-bands, npc=3). The regions attenuated by a coronagraph are masked, with a slightly larger mask on the IFS data to cover the bright signal at the edge of the coronagraphic mask. ADI data sets are displayed with one principal component to preserve features and limit geometric bias. All images use a linear colour scale and arbitrary cuts to best present all features. The disc semi-major and semi-minor axis is overlaid on the $Q_\phi$ image. Grey patches represent the FWHM of the respective non-coronagraphic PSF.}
	\label{fig:obs}
\end{figure*}

\subsection{VLT/NACO ADI data set} \label{sec:naco}

CQ~Tau was observed with VLT/NaCo \citep{Lenzen2003, Rousset2003} on the night of November 28th, 2018 in \textit{L}$^{\prime}$-band ($\lambdaup$\textsubscript{$c$}~=~3.78$\mu$m) as part of the Imaging Survey for Planets around Young stars (ISPY, \citealt{launhardt2020}) program. An Annular Groove Phase Mask (AGPM; \citealt{Mawet2005, Mawet2013}) was used to suppress stellar signal. Science frames were captured in pupil-stablised mode, enabling the application of classical ADI and PCA-ADI post-processing algorithms. The integration time was 0.35s per frame.

Since the ESO pipeline does not support coronographic imaging, we developed a new NaCo data reduction pipeline\footnote{\url{https://github.com/IainHammond/NACO\_pipeline}} utilising functions from \textsc{vip}, and used it to calibrate, pre- and post-process the data.

First, the systematically bad lower left quadrant, defined by every eighth column returning the saturated detector count, was corrected in each cube by detecting pixel columns that featured the saturated detector count and were replaced with the median of the neighbouring 5$\times$5 unaffected pixels. Cubes were automatically classified based on header keywords. For the purpose of ADI, the angle between north and the vertical axis for each science image was calculated using the starting and ending parallactic and pupil position angles. Subsequently, the rotation required to align each frame with true north was found using linear interpolation. Each derotation angle was corrected for the inherent offset from true north for NaCo (0.572$\pm0.178^{\circ}$, \citealt{launhardt2020}). 

We used a new PCA dark subtraction technique, adapted from the PCA-based sky subtraction routine implemented in \textsc{vip}, which aims to model and subtract the dark current (Appendix \ref{appendix:pca_ds}). The flat frames obtained at different airmasses were sorted based on their median pixel value and used to create a master flat frame. All science, sky, and unsaturated frames were then divided by the master flat frame. `Hot' and `cold' pixels were detected using the master flat and replaced with the median value of neighbouring 5$\times$5 unaffected pixels. Transient bad pixels (e.g., due to cosmic rays) and NaN values were also identified and sigma-filtered in 5$\times$5 pixel-boxes. NaCo's inconsistent detector integration time in the first few frames of each cube was accounted for by measuring the flux in an annulus from 3\arcsec~to the edge of the frame. The first six frames systematically featured a flux 2$\sigma$ above the median of their respective cube and were hence removed, and the remaining 94 rescaled to the median flux. A low-pass filter was applied to each unsaturated non-coronagraphic frame to determine the approximate location of the star on either of the three good detector quadrants. For each unsaturated frame, images where the star was dithered on other quadrants of the detector were used for sky subtraction to isolate the stellar PSF. A 2D Gaussian was fit to a subframe containing the median-combined PSFs, resulting in a FWHM of 4.19 pixels. 

Each science cube was paired with a sky cube captured closest in time and aligned with respect to each other using dust grains present in both cubes. We applied \textsc{vip}'s PCA-based sky subtraction algorithm with one principal component to remove the background flux, using the science cube minus the median frame of the sky cube as input and the sky cube minus the median frame of the sky cube as the PCA library.

We performed a two-step centering algorithm on the science cubes to correct for the uncertainty in location of the star behind the AGPM centre. Each science cube was first duplicated and median-combined into a single frame to enhance the speckle pattern around the star. A high-pass filter removed large-scale variations greater than the FWHM. By looping over each median-combined frame, we: \textit{i)} Performed a global centering by fitting a negative 2D Gaussian to the area around the AGPM center, and \textit{ii)} Aligned frames with respect to each other by maximising the cross-correlation of the speckle pattern expressed in logarithmic intensity. This iterative process was repeated five times to ensure the error in the centering is small. Final centering shifts were applied to individual (not median) frames of each science cube. Any frame with pixel shifts discrepant by more than 0.4 pixel with respect to the median pixel shifts along x and y was rejected. Additionally, we calculated the Pearson correlation coefficient between individual frames to the master median-combined frame in a 31$\times$31 pixel subframe around the center, rejecting frames with a correlation factor 1$\sigma$ below the median correlation factor. 

The final master cube contained 28,010 frames, temporally subsampled to 2,801 frames by taking the median of every ten frames. We post-processed the cube using \textsc{vip} implementations of median-ADI and PCA-ADI algorithms (both in full-frame and concentric annuli). We explored 1--20 principal components, with a 1$\times$FWHM inner mask and 1--3$\times$FWHM rotation threshold for PCA-ADI in concentric annuli.

\section{Results and Analysis} \label{sec:results}
Figure \ref{fig:obs} shows our final six images of CQ~Tau with VLT/NaCo, SPHERE/IRDIS and SPHERE/IFS. Our ADI images are shown with one principal component to minimise geometric biases introduced by ADI \citep{Milli2012} while the ASDI image is shown with three principal components to better emphasize each spiral feature.

\begin{figure*}
	\centering
	\includegraphics[width=\linewidth]{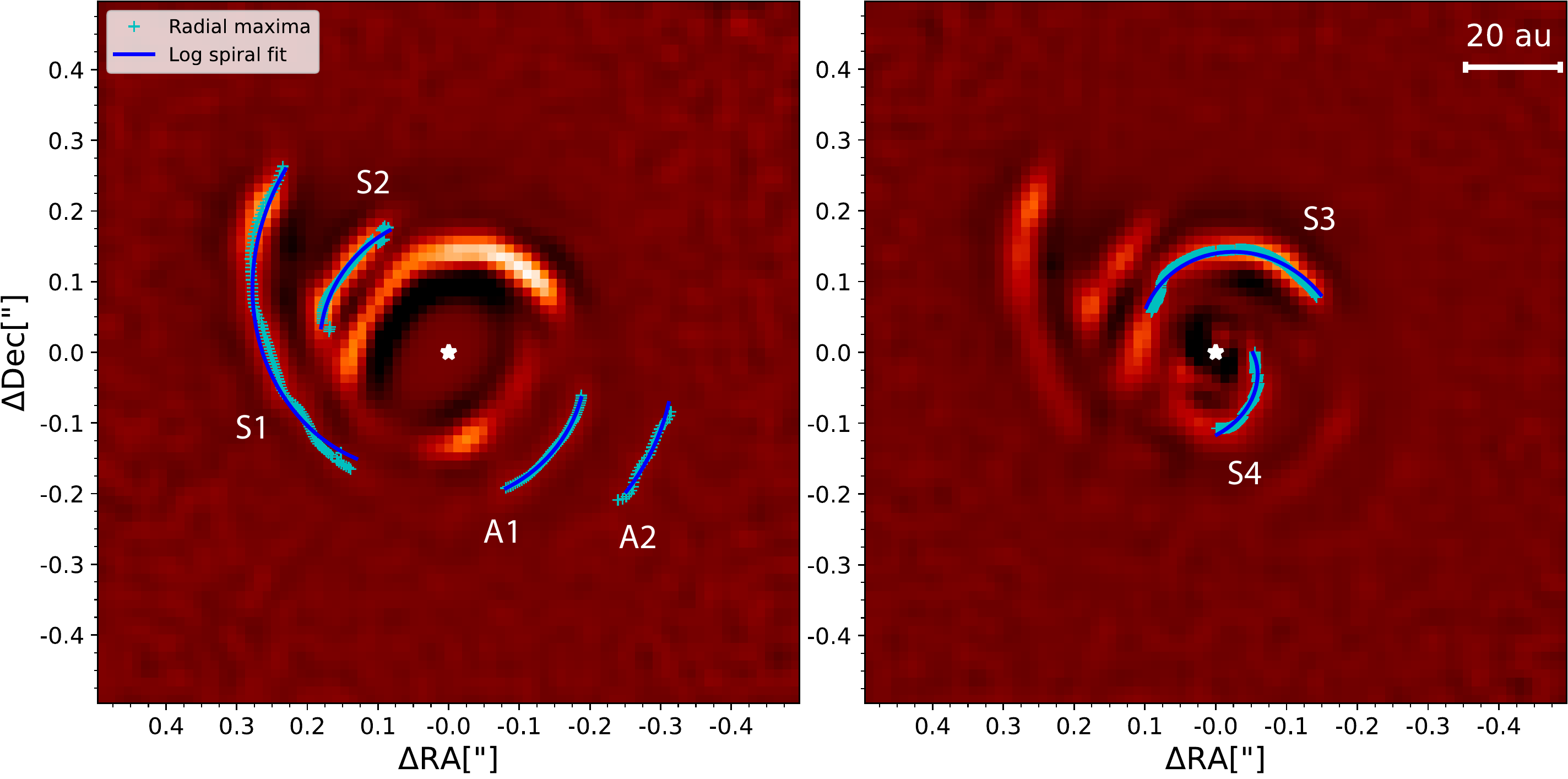} 
	\caption{Annotated versions of our coronagraphic (left) and non-coronagraphic (right) $Q_\phi$ images from Figure \ref{fig:obs}, after deprojection, $r^2$-scaling, low-pass and high-pass filtering. Best-fit logarithmic spirals for features S1--S4 and A1--A2 are presented in blue, along with their traces (cyan crosses). 
	Flux is shown in arbitrary units on a linear scale with limits chosen to best show all the features in the images.} 
	\label{fig:spiral_fit}
\end{figure*}

\begin{table}
\caption{Summary of logarithmic clock-wise spiral traces.}  
\label{tab:spiral_tab}
\begin{tabularx}{0.48\textwidth}{    
   >{\raggedright\arraybackslash}X   
   >{\centering\arraybackslash}X     
   >{\centering\arraybackslash}X     
   >{\centering\arraybackslash}X}    
\hline 
\hline
\T\B
Feature & Pitch Angle [\degr] & Root PA [\degr] & $\Delta$PA [\degr]\\
\hline
\T\B
S1          & 21.2$\pm$3.1   & 140  & 100  \\
S2          & 4.9$\pm$3.8   & 70   & 50   \\
S3          & 11.6$\pm$1.4     & 58   & 120  \\
S4          & 25.2$\pm$2.1   & 270  & 90   \\
A1          & 3.2$\pm$3.8   & 260  & 60   \\
A2          & -0.3$\pm$1.9   & 258  & 30   \\
\hline
\end{tabularx}
\end{table}
 
\begin{figure}
	\centering
	\includegraphics[width=\columnwidth]{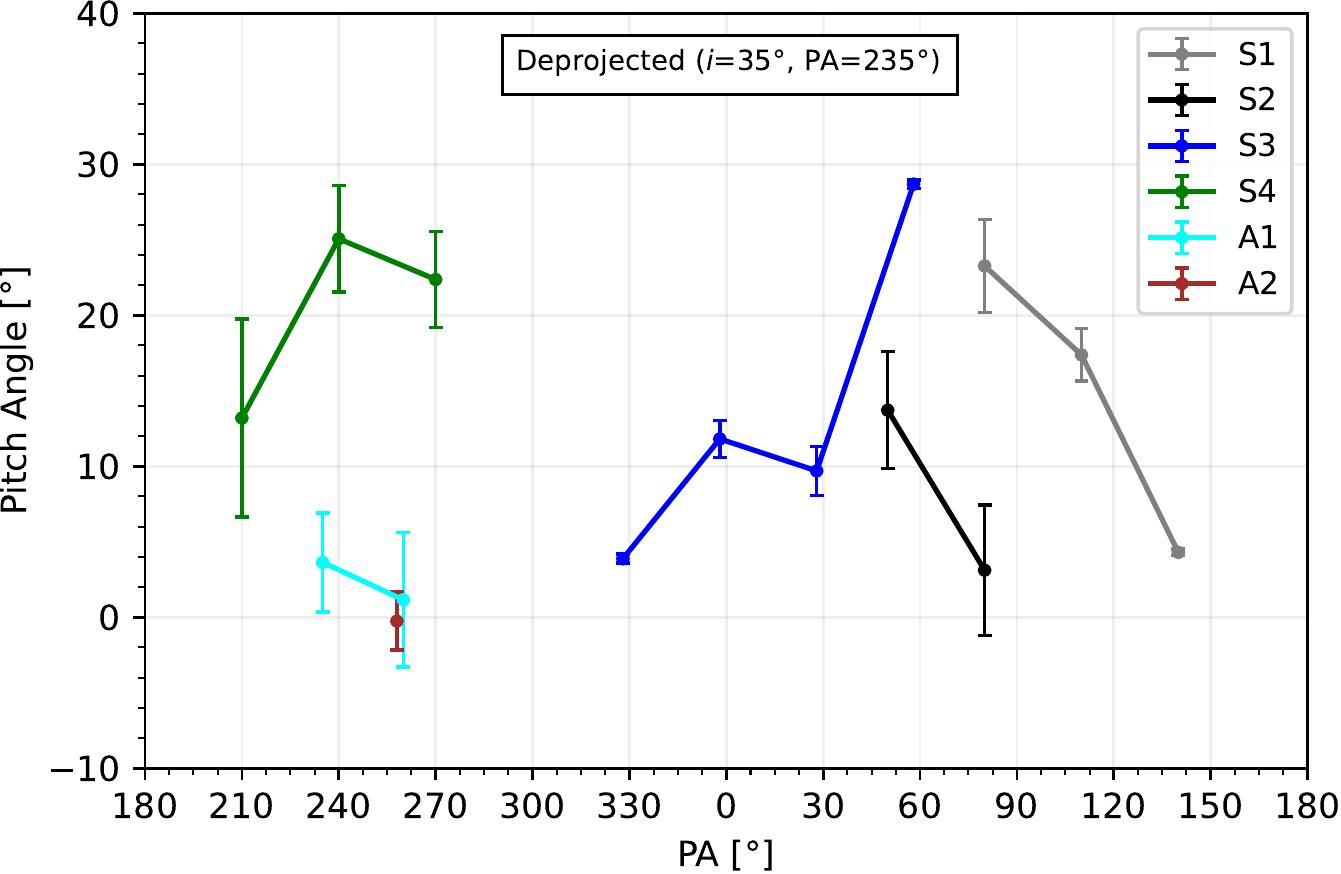} 
	\caption{Pitch angle of features S1--S4 and A1--A2. Each feature was traced using consecutive arcs subtending 30\degr~to show the evolution of the pitch angle over the entire trace. Feature A2 subtended approximately 30\degr~and thus only has one point. Uncertainties represent the difference between pitch angles using our two models for the disc scattering surface \citep{Ubeira_Gabellini2019, Bohn:2022tx} and two $Q_\phi$ images.}
	\label{fig:pitch_evo}
\end{figure}

\subsection{Spiral Arms and Shadows} \label{sec:disc_features}
 We detect spiral arms in all images. These are the most conspicuous in the $Q_\phi$ image, which is used for labelling and tracing of the spirals (Figure \ref{fig:spiral_fit}). We detect a known prominent spiral arm (we have labelled S1) to the north-east in all data sets and is robust through all tested principal components for our ADI observations. We observe a radial extent of 0\rlap{.}\arcsec2--0\rlap{.}\arcsec4 for S1, consistent with \citet{Uyama2020}. The NaCo data set features a kink in the $L^{\prime}$-band spiral emission of the prominent spiral, however we attribute this to the co-spatial, systematically bad horizontal pixels on the detector (Figure \ref{fig:pca_ds}, left) or a geometric bias of ADI and is unlikely to be a naturally occurring phenomena. A smaller tightly wound spiral (labelled S2) is found also to the north-east. 
 The non-coronagraphic $Q_{\phi}$ image best shows the presence of sub-mm grains down to several au separation, in addition to a spiral in the north (labelled S3) and emission to the south (S4) recently reported in \citet{Safonov:2022un}. There is a double-arc structure present to the SW (labelled A1 and A2) in our IRDIS ADI, PDI and IFS A/SDI images. It is unclear whether this emission has been detected in NaCo data. Features S1 and A1 appear significantly more connected (as if a prolongation of S1) in our IFS data compared to $Q_\phi$, with the exception of an azimuthally extended discontinuity in the SE in the IFS data. However, this may be 
a post-processing artefact.  

To characterise the spiral arms, we applied a spiral fitting routine to the $Q_\phi$ images. The scattering surface of the disc, derived through both the SED \citep{Ubeira_Gabellini2019} and CO rotation profiles \citep{Bohn:2022tx}, was mapped using \textsc{diskmap}'s power law function. The image was deprojected and $r^2$-scaled to account for radially decreasing stellocentric flux, knowing the \textit{Gaia} distance to the system and the inclination (35\degr) and PA (235\degr) of the outer disc. We applied a low-pass filter (Gaussian kernel equal to half the FWHM) followed by a high-pass Laplacian filter to enhance disc features and improve the automatic detection of local radial maxima - the latter accomplished using the same technique as in \citet{Casassus:2021vn}. Figure \ref{fig:spiral_fit} shows the traces detected using this method as cyan crosses. We fit a logarithmic spiral to the trace for S1--S4 and A1--A2 using \texttt{scipy.optimize}, which uses a Nelder-Mead minimization, to find the best-fitting parameters of the log spiral equation (r = \textit{a}$e^{b\theta}$) with the pitch angle defined as tan$^{-1}$(1/\textit{b}). 
As we have two functions for mapping the disc surface, contrasting pitch angles were measured for each spiral. 
For this reason, our final pitch angle uncertainty corresponds to the difference between the pitch angles of the two disc shapes and the two $Q_{\phi}$ images (S4 could only be traced in the non-coronagraphic image). The surface derived from the CO rotation profiles is close to geometrically flat in the inner $\lesssim$50 au and is unlikely to occur in nature 
however it is meaningful to explore all our available functions. Our final fits are presented in Table \ref{tab:spiral_tab} and visualised in Figure \ref{fig:spiral_fit}. Figure~\ref{fig:pitch_evo} shows the pitch angle variation found using consecutive arcs subtending 30\degr. Our errors are dominated by the uncertainty in the mapping of the disc surface.


We find pitch angles of 21.2\degr, 4.9\degr, 11.6\degr, 25.2\degr~for S1--S4. Features A1 and A2 are more consistent with circular arcs. While S2 has a shallow pitch angle, the evolution of the trace is more supportive of a spiral (Figure~\ref{fig:pitch_evo}). S1 and S2 are opening radially outward whereas the opposite is seen for S3 and S4. Our deprojected $Q_{\phi}$ images and IFS image suggest that S1 and S3 may connect through A1 (Section \ref{sec:spiral_discussion}).

The PDI and SDI data sets suggest a tentative shadow on the outer disc at two different position angles. A dimming is present at PA$\approx$140--160\degr~most clearly visible in $Q_\phi$, also reported in $^{12}$CO, $^{13}$CO and C$^{18}$O $J$=2--1 peak brightness temperature (\citealt{Wolfer2020}).

\subsection{Companion Detection Limits} \label{sec:detection_limits}

\begin{figure}
	\centering
	\includegraphics[width=\columnwidth]{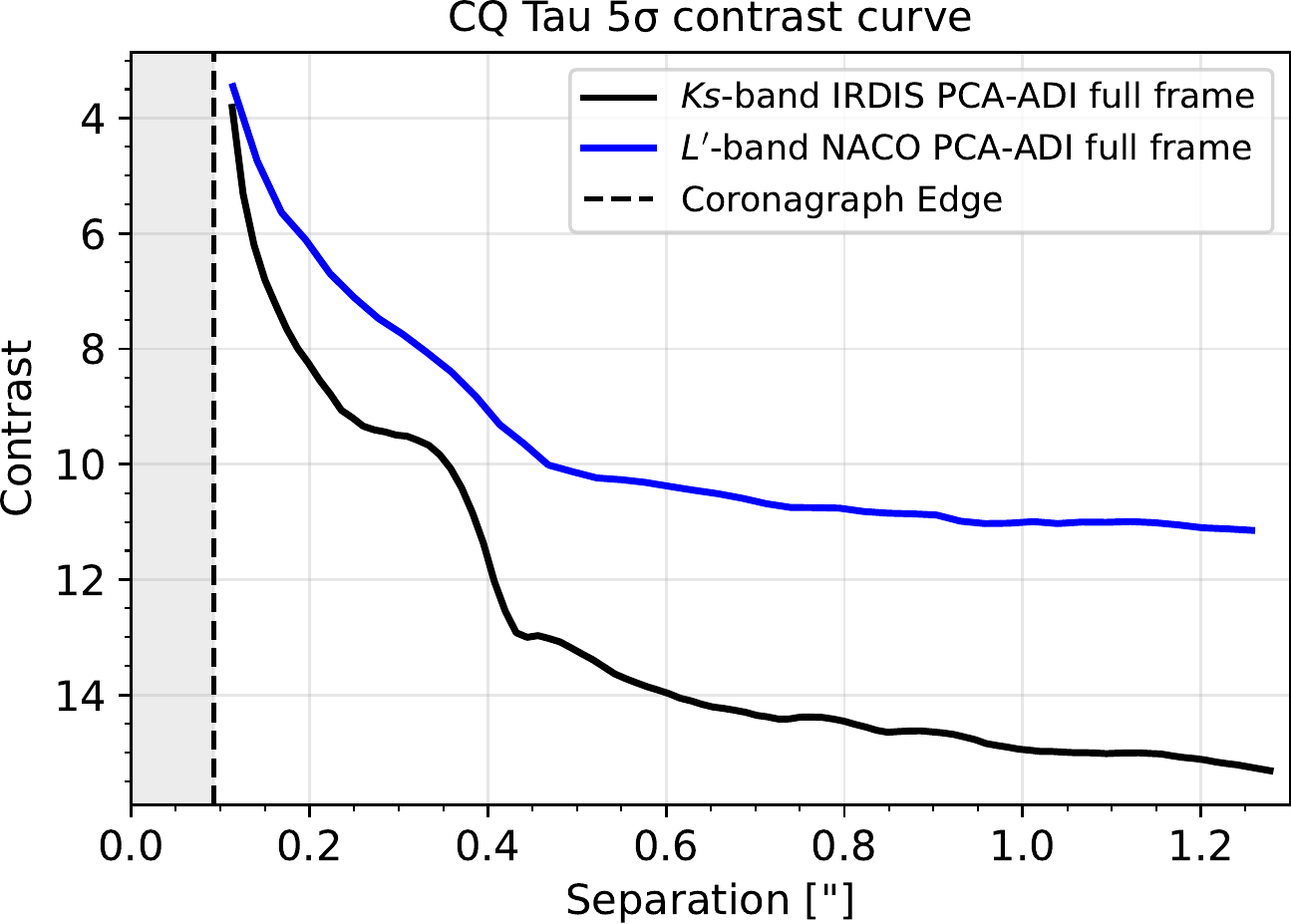}
	\caption{5$\sigma$ contrast (in magnitudes) with respect to the star computed for our NaCo and IRDIS ADI data sets. The contrast for each radial separation was selected using the optimal contrast achieved by varying the number of principal components in Section \ref{sec:detection_limits}. The ALC2 (IRDIS) and AGPM (NaCo) coronagraphs hide the inner 14 and 16 au respectively.} 
	\label{fig:contrast}
\end{figure}

\begin{figure}
	\centering
	\includegraphics[width=\columnwidth]{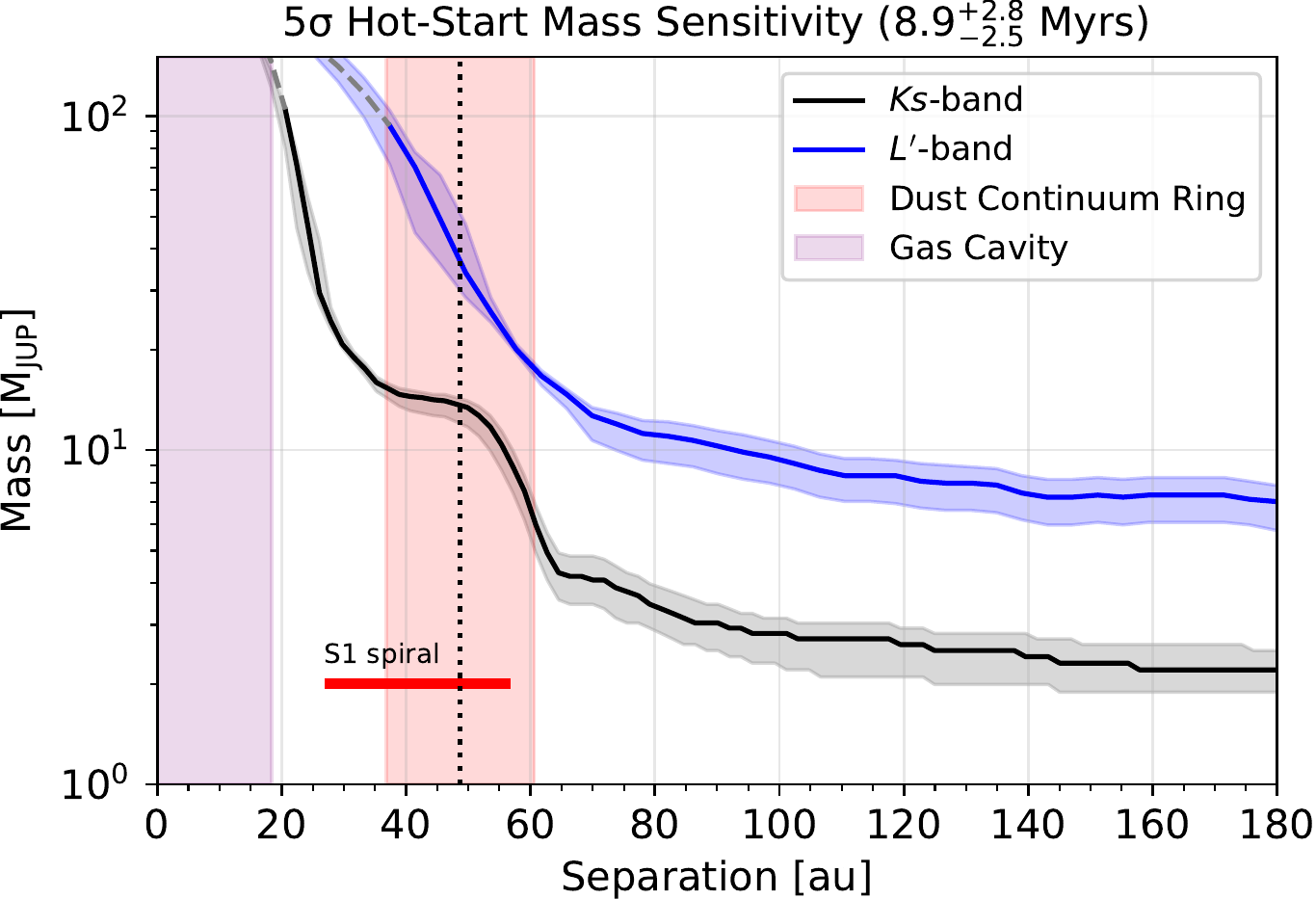}
	\caption{Mass sensitivity curve for \textit{Ks}-band and \textit{L}$^{\prime}$-band full frame PCA-ADI reductions 
	determined with the hot-start COND evolution models \citep{Baraffe:2003uy} at 8.9 Myrs. Shading around each curve corresponds to the age uncertainty. 
	The dashed line represents an extrapolation of the models beyond 104~M$_{\rm Jup}$. Red line gives the S1 spiral extent. Purple shading indicates the gas cavity. Vertical dotted line indicates the center of the 1.3mm dust continuum ring (\citealt{Ubeira_Gabellini2019}, \citealt{Wolfer2020}).} 
	\label{fig:mass_sensitivity}
\end{figure}

Apart from the point source at $\sim$2\rlap{.}\arcsec2 separation (327 au, or $\sim$250~au from the edge of the dust continuum ring)  that we determined to be a background object (Appendix~\ref{sec:astrometry}), our final post-processed images did not contain any reliable point sources in \textit{Ks} or \textit{L}$^{\prime}$-band over a wide range of tested principal components (1 to 20). Figure \ref{fig:contrast} shows the 5$\sigma$ contrast curve and Figure \ref{fig:mass_sensitivity} the mass sensitivity limits for our two ADI data sets without subtracting the S1 spiral. The 3$\sigma$ sensitivity at 0\rlap{.}\arcsec3 and 0\rlap{.}\arcsec6 was first determined using \textsc{vip}'s contrast curve function. The purpose of this step was to infer the approximate flux recovered of synthetic planets with S/N of ten using full frame PCA-ADI to 20 principal components. 
Knowing the approximate flux, we created six ADI cubes each containing a synthetic companion at 0\rlap{.}\arcsec3 and 0\rlap{.}\arcsec6, separated in PA by 60$^{\circ}$ forming a `spiral' of uniquely positioned synthetic companions in each cube. The cubes were post-processed with full frame and annular PCA-ADI techniques to 1--20 principal components, the latter using 3$\times$FWHM pixel annuli and a rotation threshold of 1--3$\times$FWHM for the inner and outer annuli. We compute the S/N for the synthetic companions in each cube to determine the optimal number of principal components, at each separation, ignoring negative side lobes extending azimuthally from the companions and accounting for small sample statistics as in \citet{Mawet2014}. Our contrast at each separation was then the optimal contrast achieved with our range of principal components. The final 5$\sigma$ contrast curve (Figure \ref{fig:contrast}) presents the optimal contrast obtained with our post-processing techniques and provides deeper contrast than previously reported \citep{Asensio-Torres:2021vr} in the noise-limited regime.

We computed the expected absolute magnitude of 0.5--104~M$_{\rm Jup}$ giant planets and brown dwarfs using hot-start COND evolution models \citep{Baraffe:2003uy} with a linear interpolation between the 5 and 10 Myr magnitudes to 8.9 Myrs. We use an apparent magnitude of 6.173 (2MASS, \citealt{2003yCat.2246....0C}) and 5.043 (AllWISE, \citealt{2014yCat.2328....0C}) in \textit{Ks} and \textit{L}$^{\prime}$-band, respectively, and the updated \textit{Gaia} distance (Table \ref{tab:CQ_Tau_prop}) for CQ~Tau to convert our contrast curve into absolute magnitude limits. Based on the analysis of \citet{Sanchis:2020wc}, we can exclude a significant effect of disc extinction at these distances from the star and these values of the potential planet mass. Our \textit{Ks}-band mass sensitivity curve rules out giant planets above 4~M$_{\rm Jup}$ outside the spiral region 
and 2--3~M$_{\rm Jup}$ beyond 100 au to 5$\sigma$ confidence (Figure \ref{fig:mass_sensitivity}). The end of the NIR spiral region is characterised by a sharp drop in the contrast curve occurring at $\sim$60 au. At that location, we enter the noise-limited regime and our contrast is no longer biased by the bright spirals. This curve does not rule out a brown-dwarf companion interior to the sub-mm ring, or a stellar companion in the gas cavity. The discrepancy between the contrast achieved with NaCo (\textit{L}$^{\prime}$-band) and SPHERE (\textit{Ks}-band) is primarily due to the higher sky background in \textit{L}$^{\prime}$-band.

Uncertainty in the age of CQ~Tau (Table \ref{tab:CQ_Tau_prop}) will influence our companion magnitudes (brighter at younger ages). For this reason we re-computed our mass sensitivity using our constraints on the age of the system (6.3--11.7 Myrs), shown as the shaded region in Figure \ref{fig:mass_sensitivity}. Since temporal subsampling of our NaCo data can reduce the S/N of a companion at large (>0\farcs5) separation from the host star \citep{GomezGonzalez2017}, we also computed the \textit{L}$^{\prime}$-band mass sensitivity without binning with 20 principal components using both PCA-ADI flavours. No substantial change in contrast was detected, suggesting a median binning factor of ten frames was not substantial enough to smear a candidate companion's signal nor decorrelate the PSF. One can also infer that a single tested principal component will not provide optimal contrast, but instead sampling over a range of components will give the deepest contrast possible at each radial separation (as conducted in Figure \ref{fig:mass_sensitivity}).

\section{Discussion} \label{Discussion}

\subsection{Where is the spiral-inducing companion?}\label{sec:spiral_discussion}
Our non-detection of a wide orbit companion inducing the spiral arms in CQ~Tau is consistent with the cases of HD~135344B and MWC~758, where a $\sim$3~M$_{\rm Jup}$ companion exterior to the spirals should have been detected to 5$\sigma$ confidence \citep{Maire2017, Reggiani2018, Boccaletti:2021tl}. Taking into account the multi-epoch astrometry of the point-source detected in Figure \ref{fig:background}, we rule out the possibility this source is bound to the system and thus is not dynamically influencing the disc. While our \textit{L}$^{\prime}$-band images did not reach equivalent contrast to our \textit{Ks}-band data, the Keck/NIRC2 ADI data presented in \citet{Uyama2020} set companion mass limits of $\sim$5~M$_{\rm Jup}$ outside the prominent spiral to 0\rlap{.}\arcsec9. The minimum mass required (on the order of several Jupiter masses) to excite such high-amplitude spiral arms, based on simulations \citep{Dong2015a, Zhu2015a, Juhasz2015, Dong2017b}, is likely to be detectable in our IRDIS and NaCo data sets assuming a ``hot start" (Figure \ref{fig:mass_sensitivity}). Furthermore, CQ~Tau does not feature a two-arm spiral pattern separated by $\sim$180\degr~expected for high-mass external companion \citep{Fung2015}, although we recognise this conclusion could be influenced by illumination effects (Section \ref{sec:shadow_discussion}). 
However, the recovered pitch angle for S3 is within expectations for inner spiral arms excited by a several Jupiter mass perturber \citep[$\sim$10--15\degr,][]{Dong2015a}. The pitch angle of the S1 feature in particular (34$\pm$2\degr~in \citealt{Uyama2020}, 21.2$\pm$3.1\degr~in this work due to our deprojection of the disc, Section \ref{sec:disc_features}) would be more favourably explained by an external perturber, however can also form if an inner companion is on an eccentric orbit.

\begin{figure}
	\centering
	\includegraphics[width=\columnwidth]{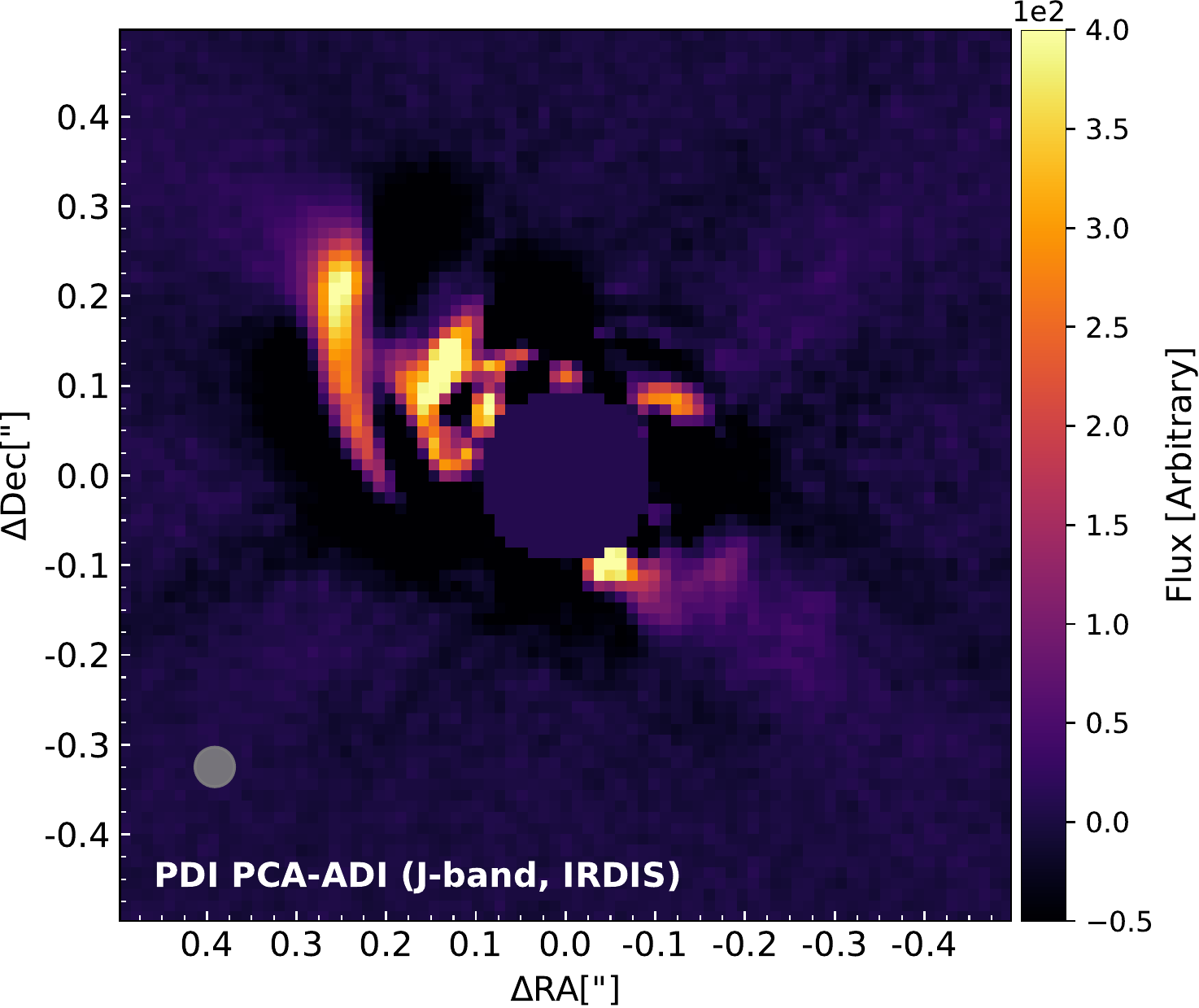} 
	\caption{The result of applying PCA-ADI to our coronagraphic $Q_\phi$ image with one principal component to test the effect of PCA-ADI on azimuthally extended signals. Flux is shown in arbitrary units on a linear scale with limits chosen to best show all the features in the image. Extended disc signals (i.e., spirals) have been strongly filtered. }
	\label{fig:PDI_PCA-ADI}
\end{figure}

The mass sensitivity estimates are dependent upon a well constrained age, distance and apparent magnitude of the system. 
If the system is older than the 8.9 Myrs used to compute the COND model magnitudes in Figure \ref{fig:mass_sensitivity}, our 5$\sigma$ detection limit will not reach as low mass and a companion may not be detected, and vice versa. We can however use the NICEST extinction map (\citealt{Juvela:2016th}) based off the Two Micron All-Sky Survey to make the observation that CQ~Tau is not spatially associated with the Taurus star forming region, as defined in \citet{Lombardi:2010vf}. Therefore, CQ~Tau may be older than the 1--3 Myr Taurus cloud complex, in agreement with the aforementioned age range. A further consideration is the choice of post-formation luminosity of a companion, which can lead to significant magnitude discrepancies at the same age \citep{Spiegel2012}. High-contrast imaging is less supportive of cold-start models associated with core-accretion \citep{Wang:2018uz, Marleau:2019wg} and favour warm- or hot-start initial conditions (although detection bias favours brighter, hot-start planets). The use of the hottest, most luminous models AMES-DUSTY \citep{Allard:2001vc} would make our \textit{Ks}-band images sensitive to lower mass planets. On the other hand, Bern EXoplanet cooling tracks with COND atmospheres and warm-start initial conditions (\citealt{Marleau:2019vx, Vigan:2021ww} and recently applied to SPHERE data in \citealt{Asensio-Torres:2021vr}) indicate a planet with mass on the order of $\sim$8~M$_{\rm Jup}$ may not be detected in our \textit{Ks}-band images immediately outside the S1 spiral. CQ~Tau is a relatively old disc allowing time for a slower planet formation, which may be at play. We therefore cannot entirely rule out an external perturber with our observations.

Searching for a planet or companion interior to the spirals is a logical next step. While \citet{Ubeira_Gabellini2019} successfully fit the depletion of dust and gas tracers due to a 6--9~M$_{\rm Jup}$ planet at 20 au with hydrodynamical models, spirals and non-axisymmetric features were not discussed. It is therefore natural to question whether an inner perturber could explain such features.

Our ADI data in Figure \ref{fig:obs} features several point-like sources and filtered spiral signals. However PCA-ADI introduces geometric biases to azimuthally extended disc signals, transforming them into point-like sources. We therefore tested the application of PCA-ADI on our PDI data set. A fake ADI cube was built from rotating the $Q_{\phi}$ image using the same derotation angles as the IRDIS ADI observation, and subsequently post-processed with PCA-ADI to one principal component. The image, which we have labelled PDI PCA-ADI (Figure \ref{fig:PDI_PCA-ADI}), features the prominent single spiral arm S1 present in all our data sets, filtered versions of S2, S3, S4, A1 and potentially A2. The azimuthally extended emission from S2 and S3 has been most strongly filtered, such that the residual signals are analogous to those observed in our ADI data. Therefore, the distortion (or absence entirely) of these structures in ADI images is rather ADI filtering. 

\begin{figure}
	\centering
	\includegraphics[width=\columnwidth]{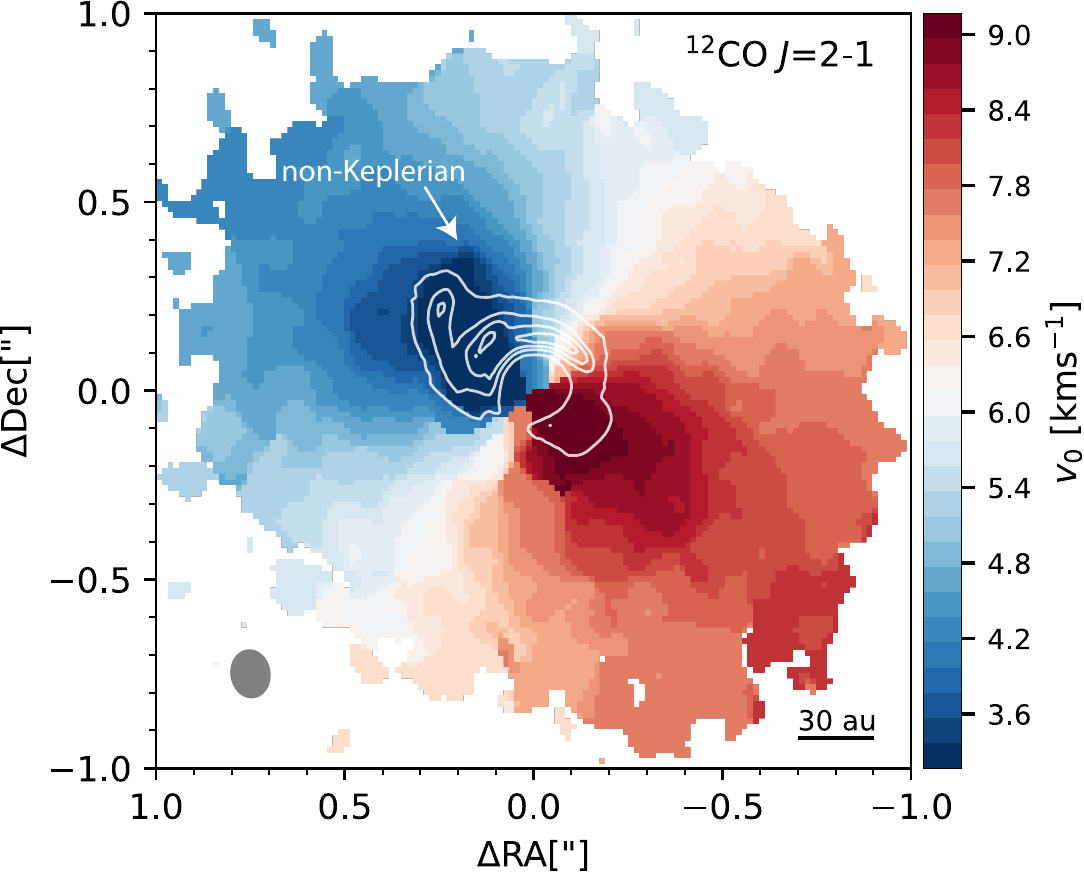} 
	\caption{$Q_\phi$ $r^{2}$-scaled contours overlaid on $^{12}$CO \textit{J}=2--1 rotation map, $v_{0}$, produced with the quadratic method 
	in \textsc{bettermoments} \citep{Teague:2018wr}. Regions with velocities not between zero to 12 kms$^{-1}$ are masked. The arrow highlights the non-Keplerian velocity signature observed in the vicinity of the NIR spiral. CO data originally presented in \citet{Wolfer2020} (data availability within).}
	\label{fig:12co_panel}
\end{figure}

Our data sets suggest a single large-scale spiral arm (S1) in CQ~Tau, accompanied by the existence of three smaller spirals (S2, S3 and S4) and a double-arc (A1 and A2). Figure~\ref{fig:12co_panel} shows the S1 spiral arm connects into a large-scale velocity perturbation in the $^{12}$CO data, suggesting that this is part of a broader spiral structure \citep{Wolfer2020}. A single large-scale or far-reaching spiral is not unusual, as seen in polarised light images of HD 100546, V1247 Ori and HD 142527 (\citealt{de-Boer:2021tv, Ohta2016, Avenhaus:2014vu}, respectively). In the case of HD 142527, all disc features have been qualitatively reproduced by the simulated interaction of the disc and an inclined, eccentric binary companion \citep{Price:2018tu}. The presence of a planetary-mass companion or close-in binary remains ambiguous in the case of HD 100546 and V1247 Ori, however candidates have been proposed \citep{Sissa:2018te,Fedele:2021uu,Kraus2017}. Similarly, the binary system HD 34700 features a spiral of greater radial extent and pitch angle relative to a series of smaller spirals present in scattered light images \citep{Monnier:2019vo}. Our mass limits support an analogous scenario, although CQ~Tau features a smaller dust cavity compared to HD 142527 ($\sim$90 au) and lacks the extreme dust asymmetry of V1247 Ori, HD 142527 and HD 34700 (\citealt{Ohta2016,Perez2014,Benac:2020wi}, respectively). 

Multi-epoch $I_{c}$-band ($\lambdaup$\textsubscript{eff} = 0.8$\mu$m) differential speckle polarimetry (DSP) observations by \citet{Safonov:2022un} spanning 2015--2021 detected a northern and southern spiral spanning 0\farcs1--0\farcs15 from the star (which could account for some of the stellar variability, \citealt{Dodin:2021tj}). The morphology recovered in our non-coronagraphic image at the same PA and separation is consistent with the DSP image, providing an independent confirmation this feature is real. While the authors suggested the northern spiral was S2, the feature is rather spatially associated with S3. A similar spiral to the south (PA$\approx$200\degr) is associated with emission best seen in our non-coronagraphic $Q_\phi$ image and is partially visible in our coronagraphic data sets (Figure \ref{fig:obs}). \citet{Safonov:2022un} estimated the pattern speed of S3 to be $-0.2\pm1.1$\degr/yr, compared to $\sim$4.8\degr/yr if induced by a companion at 20 au in \citet{Ubeira_Gabellini2019}. If caused by a closer-in companion at shorter separation, one would expect a higher magnitude pattern speed of the spiral corresponding to the companion's Keplerian velocity. The authors suggest S3 could instead be tracing the edge of an outer dust disc however our measured pitch angle is less supportive of this. This analysis would benefit from independent confirmation with multi-epoch SPHERE/IRDIS and Subaru/AO188+HiCIAO polarimetric observations as only 135 days passed between our PDI data sets.

Figure \ref{fig:spiral_fit} and Table \ref{tab:spiral_tab} show small ($<$4\degr) pitch angles for A1 and A2. Our images and spiral traces (Figure \ref{fig:pitch_evo}) suggest that A1 could be connected to S1 and S3, forming part of the same spiral arm that has been visually segmented due to the illumination of the disc. If so, the trace of this feature would represent a pitch angle reversal which can only be explained with an inner companion on an eccentric orbit also predicted for MWC~758 (Figure 3 in \citealt{Calcino2020}). Such a perturber is favorable given our detection limits. In the case of A2, the arc-like appearance 
is reminiscent of HD~100453 \citep{Benisty2017} and MWC~758 \citep{Benisty2015,Reggiani2018}. In both systems, this was attributed to the outer edge of the scattering surface on the bottom disc and located near the disc semi-minor axis. However in MWC~758, the arc is coincident with a dust trap (see Figure 1, \citealt{Dong:2018um}) which could suggest scattered light is probing the edge of a large dust concentration. We speculate the A2 feature may also be probing the edge of a large dust collection in CQ~Tau considering its location along the disc semi-major (not minor) axis, near-zero pitch angle and proximity to the southern 1.3 mm continuum peak.


\subsection{Are the shadows real? If so, what is causing them?} \label{sec:shadow_discussion}
A drop in flux observed both towards the SE and west (this is most clear in $Q_\phi$ and when enhanced in Figure \ref{fig:spiral_fit}) suggests shadowing onto the outer disc from the proposed inner disc (r$\approx$0.2 au, \textit{i}$\approx$48\degr, PA$\approx$106\degr~in \citealt{Eisner:2004wd} and illustrated in \citealt{Safonov:2022un}, \textit{i}$\approx$29\degr, PA$\approx$140\degr~in \citealt{Bohn:2022tx}). A drop in temperature co-located with the reduction in brightness was detected in CO isotopologues \citep{Wolfer2020} to the southeast and northwest. The authors cite beam dilution as a possible mechanism due to its proximity to the disc semi-minor axis, however the re-detection of the southeast shadow in scattered light suggests this may be gas cooling in the presence of the shadow (e.g., HD~142527, \citealt{Casassus:2015tm,Casassus:2019ve}). The large-scale scattered-light surface brightness asymmetry between the shadows 
closely resembles HD~143006, where two dark lanes give way to an underbrightness in half of the disc \citep{Benisty:2018uw}. A warped disc model due to an inclined binary companion successfully reproduced the shadow, with a 10~M$_{\rm Jup}$ circumbinary planet producing the dust cavity \citep{Ballabio:2021wt}. 

\begin{figure}
	\centering
	\includegraphics[width=\columnwidth]{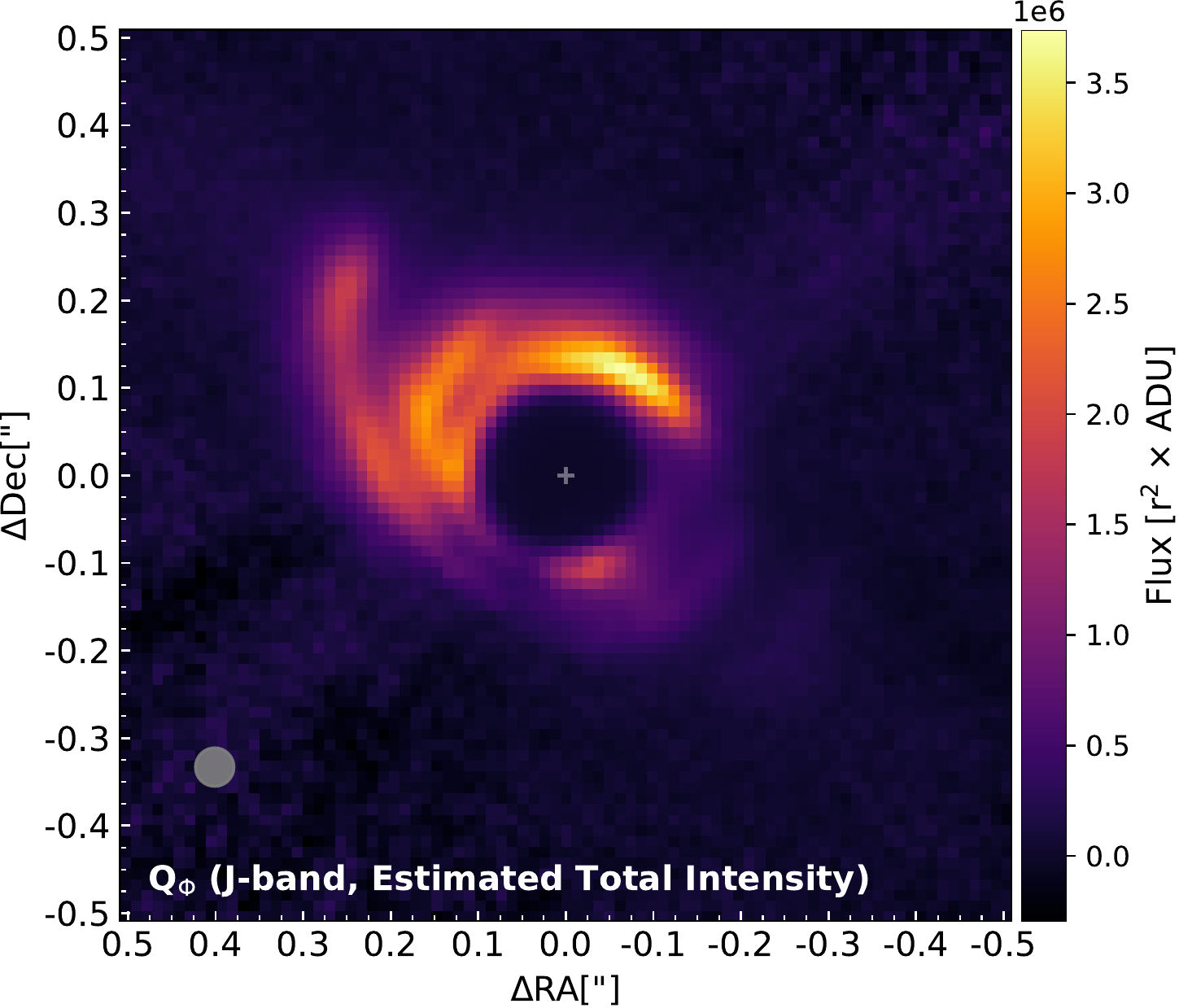}
	\caption{Estimated total intensity of the $r^{2}$-scaled coronagraphic $Q_\phi$ image after correcting for the degree of polarisation with \textsc{diskmap}. Flux is displayed on a linear scale with an intensity cut from the 0.1--100 percentiles. } 
	\label{fig:total_intensity}
\end{figure}

To test whether the shadows are real or a polarisation phase function artefact, we show the estimated total intensity image in Figure \ref{fig:total_intensity}. As the shadows and large-scale brightness asymmetry still appear in the total intensity image, this implies the dimming is unlikely to be caused by the scattering properties of the dust grains in the disc.


Our observations are currently most supportive of a fairly axially symmetric shadow expected for a misaligned inner disc \citep{Whitney:2013ve}, due to the presence of a massive companion in the cavity \citep{Owen2017}. A recent study of the inner disc geometry using VLTI/GRAVITY \citep{Bohn:2022tx} 
found its inclination of 29$\pm$3\degr~was a small deviation from that of the outer disc (32$\pm$1\degr) determined with $^{12}$CO and $^{13}$CO line data, however the position angle of the inner and outer disc major axis varied significantly (140$^{+7}_{-9}$\degr~compared to 234$\pm$1\degr, respectively). The expected shadows cast onto the disc agree well with the darker regions and large-scale asymmetry. The discrepancy between the inner disc orientation presented in \citet{Eisner:2004wd} could be due to its precession over the two epochs \citep[e.g.,][]{Bate:2000ub}. Multi-epoch observations would further constrain the presence and movement of the inner disc. As the proposed inner disc precesses, the shadows should appear to `rock' back and forth in azimuth over time \citep{Nealon:2020tl}. Irregularly changing patterns could indicate a variety of formation mechanisms (e.g., HD 135344B, \citealt{Stolker2017}) such as super-Keplerian transient shadows projected by an inner companion with a circumplanetary disc (e.g., GG~Tau, \citealt{Itoh:2014vp, Keppler:2020vi}) or shadowing from the spiral itself (e.g., HD~34700, \citealt{Montesinos:2016tf}). 

\subsection{What is causing the UX Ori variability?} \label{sec:binarity}

UX~Orionis stars are characterised by photometric variations of several magnitudes in the optical that coincide with increased extinction and polarisation, hypothesized as being due to the disturbance of circumstellar dust by a companion or binary \citep{Grinin:1994tl, Grinin:2008}. 
\citet{Shakhovskoj:2005ux} reported a 7800 day ($\sim$21.35 year) cycle in the light curve data for CQ~Tau, with a secondary harmonic of 3900 days ($\sim$10.67 years). When subtracted from the periodogram, a 1020 day ($\sim$2.79 year) cycle is revealed (a cycle ratio close to that of UX~Ori itself). American Association of Variable Star Observers data consisting of over 11,000 observations since 1946 
presents the same approximate ten year period. If the 21.35 year cycle were to be caused by a companion orbiting the star with a combined mass of 1.5~M$_\odot$, this equates to an orbital separation of 8.8 au. Interestingly, when following the theoretical evolution of co-planar binaries in protoplanetary discs, the semi-major axis of the cavity reaches a$_{\rm cav}$ $\approx$ 3.5$a_{\rm bin}$ \citep{Artymowicz1994,Hirsh2020,Ragusa:2020vl}. In the case of CQ~Tau with a gas cavity upper limit of 25 au, we also arrive at an orbital separation of $\sim$8 au. However, models by \citet{Demidova:2010tt} suggest the shorter cycle in UX~Ori stars should correspond to a binary orbit ($\sim$2.2~au using the shorter cycle for CQ~Tau) and its material streams, while the secondary cycle is caused by spiral waves \citep{Sotnikova:2007uj}. 
Spectra from the Echelle SpectroPolarimetric Device for the Observation of Stars (ESPaDOnS) strongly supports the presence of a transient inhomogeneous dust cloud, comparable to the stellar radius, with crossing times corresponding to a distance of at least 0.9 au \citep{Dodin:2021tj}. 

\textit{Gaia} DR3 finds a poor fit to the single-star model \citep{Lindegren:2021} indicated by the 
astrometric excess noise of 0.45mas with a significance of 274, and dimensionless re-normalised unit weight error (RUWE) $\simeq$ 3.7. The RUWE (ideally = 1 for a good fit to the single star model), which is sensitive to unresolved binaries, is tentatively indicative of an unresolved object. 
The 15$\pm$2 km s$^{-1}$ radial velocity (LSRK) of the star is compatible with an M--dwarf companion on a $\sim$2 au orbit as predicted by \citet{Demidova:2010tt}, although measurements will be biased by accretion streams and the precession of, and obscuration by, the inclined inner disc. The presence of a circumstellar disc can also result in inflated RUWE values, prompting a higher cutoff for binaries in the presence of disc material. \citet{Fitton:2022uk} proposed such a cutoff at RUWE~=~2.5, however CQ~Tau exceeds this value.

For comparison, these astrometric errors are larger than for PDS~70 ($d$ = 112 pc) -- a transition disc with two directly imaged protoplanets (e.g., \citealt{Keppler2018, Benisty:2021uj}) -- and smaller than those found for the stellar binary system GG~Tau~A (projected separation $\sim$38 au, RUWE $\simeq$ 6.4) at a comparable distance as CQ~Tau. Most interestingly, the astrometric errors for CoKu~Tau/4 ($d$ = $\sim$150 pc, astrometric excess noise of 0.8mas with a significance of 293, RUWE $\simeq$ 5.9), a near-equal mass binary separated by 8~au \citep{Ireland:2008vh}, are similar to those for CQ~Tau. 
A binary companion would account for both the UX Ori-type variability and \textit{Gaia} measurements, however it is unclear at this time whether its separation would be 2~au or 8~au.




\section{Conclusion} \label{Conclusion}
We combined all available ESO high-contrast imaging data of CQ~Tau to investigate the hypothesis of a companion dynamically influencing the disc. In addition, we developed a new, open-source NaCo data reduction pipeline for reducing any coronagraphic NaCo data set with modern techniques, and implemented a PCA-based dark subtraction algorithm. Our key conclusions are as follows:

\begin{enumerate}

\item We confirm the detection of an extended (0\rlap{.}\arcsec2--0\rlap{.}\arcsec4, Figure \ref{fig:obs}) prominent spiral arm \citep{Uyama2020}, bright in \textit{Y}--\textit{L}$^{\prime}$-band wavelengths co-spatial with the dust continuum peak and large-scale velocity perturbation (Figure \ref{fig:12co_panel}), in addition to detecting three smaller spirals. Logarithmic spiral fitting resulted in a pitch angle of 21.2$\pm$3.1\degr, 4.9$\pm$3.8\degr, 11.6$\pm$1.4\degr~and 25.2$\pm$2.1\degr, respectively. Two arcs are detected, with one segment likely joining a larger spiral arm that appears discontinuous due to the illumination of the disc, while another is co-incident with the dust continuum (Section \ref{sec:spiral_discussion}).
\item Companion detection limits appear to rule out a hot-start companion above 4~M$_{\rm Jup}$ outside the spiral region in \textit{Ks}-band and 2--3~M$_{\rm Jup}$ beyond 100 au in to 5$\sigma$ confidence (Figure \ref{fig:mass_sensitivity}). We detected NIR sources between $\sim$0\rlap{.}\arcsec15 and $\sim$0\rlap{.}\arcsec4 in our ADI data sets, however we can not distinguish them from filtered disc features or residual speckles (Figure \ref{fig:PDI_PCA-ADI}).
\item Multi-epoch astrometry strongly suggests the point source at $\sim$2\rlap{.}\arcsec2 separation is not bound to the system (nor a stellar flyby candidate), and CQ~Tau should not be considered a binary based off this object (Appendix \ref{sec:astrometry}).
\item Disc shadowing and illumination asymmetry is detected in our PDI data (Figure \ref{fig:obs}). Two dark lanes are close to axially symmetric and likely caused by the misaligned inner disc best compared to HD~143006. Multi-epoch observations would constrain their change in azimuth.
\item Our results (points \textit{i--iv}), considered together with ALMA, GRAVITY and differential speckle polarimetry data \citep{Ubeira_Gabellini2019,Wolfer2020,Bohn:2022tx,Safonov:2022un} support the hypothesis of an inner companion on an inclined orbit forming a misaligned inner disc (Section \ref{Discussion}). Such a companion could be on an eccentric orbit driving the spiral pattern \citep{Calcino2020, Debras:2021us} and high \textit{Gaia} RUWE, although dedicated models are required to re-
produce the large pitch angle of the prominent spiral.

\end{enumerate}


We suggest new SPH simulations that extend the work of \citet{Demidova:2010tt,Demidova:2010um}, \citet{Demidova:2015tp} and \citet{Calcino2020} to replicate the large scale spiral structure. Upcoming \textit{Gaia} astrometry and binary orbital fits could constrain the orbit of an unseen companion within the gas cavity. The narrow separation of the proposed companion makes it favourable for observations with the Keck Planet Imager and Characterizer in Vortex Fiber Nulling mode. 
Follow-up observations with upcoming infrared high-contrast instruments on the Extremely Large Telescope could provide the contrast and resolution required to fully comprehend the mechanisms driving the substructures in CQ~Tau.

\section*{Acknowledgments} 

We thank the referee for their useful suggestions. IH acknowledges the support of a Research Training Program scholarship from the Australian government. IH, VC, DP and CP acknowledge funding from the Australian Research Council via DP180104235 and FT170100040. VC also acknowledges funding from the Belgian F.R.S.-FNRS for financial support through a postdoctoral researcher fellowship. CT, GL, LT and MGUG acknowledge funding from the European Union’s Horizon 2020 research and innovation programme under the Marie Skłodowska-Curie grant agreement No 823823 (RISE DUSTBUSTERS project). LT acknowledges support from the Italian Ministero dell’Istruzione, Università e Ricerca through the grant Progetti Premiali 2012-iALMA (CUP C52I13000140001), from the Deutsche Forschungsgemeinschaft (DFG, German Research Foundation) - Ref no. 325594231 FOR 2634/2 TE 1024/2-1, from the DFG Cluster of Excellence Origins (www.origins-cluster.de), and from the European Research Council (ERC) via the ERC Synergy Grant ECOGAL (grant 855130). MB acknowledges funding from the European Research Council (ERC) under the European Union's Horizon 2020 research and innovation programme (grant PROTOPLANETS No. 101002188). DF acknowledges the support of the Italian National Institute of Astrophysics (INAF) through the INAF Mainstream projects ARIEL and the ``Astrochemical Link between Circumstellar Disks and Planets", ``Protoplanetary Disks Seen through the Eyes of New- generation Instruments" and by the PRIN-INAF 2019 Planetary Systems At Early Ages (PLATEA). This work has made use of the Multi-modal Australian ScienceS Imaging and Visualisation Environment (MASSIVE) (\url{www.massive.org.au}). This publication makes use of data products from the Wide-field Infrared Survey Explorer, which is a joint project of the University of California, Los Angeles, and the Jet Propulsion Laboratory/California Institute of Technology, funded by the National Aeronautics and Space Administration (NASA). This work has made use of data from the European Space Agency (ESA) mission \textit{Gaia}. 
Our data reduction and analysis made use of the \textsc{python} programming language and the \textsc{astropy} \citep{Astropy-Collaboration:2013uv,Astropy-Collaboration:2018vm}, \textsc{numpy} \citep{Harris:2020ti}, \textsc{matplotlib} \citep{Hunter:2007ux}, \textsc{photutils} \citep{Bradley:2019vy}, \textsc{scikit-image} \citep{scikit-image}, \textsc{scikit-learn} \citep{scikit-learn}, and \textsc{scipy} \citep{Virtanen:2020uj} packages.

\section*{Data Availability}
The reduced images will be made available on the Strasbourg astronomical Data Center (CDS) and through Monash University's \textit{Bridges} research repository via \href{https://doi.org/10.26180/20372814}{doi:10.26180/20372814}. Data reduction algorithms used in this article are available within the article.
\bibliographystyle{mnras}
\bibliography{CQ_Tau.bib} 





\appendix
\section{PCA Dark Subtraction with NaCo} \label{appendix:pca_ds}

\begin{figure}
	\centering
	\includegraphics[width=\columnwidth]{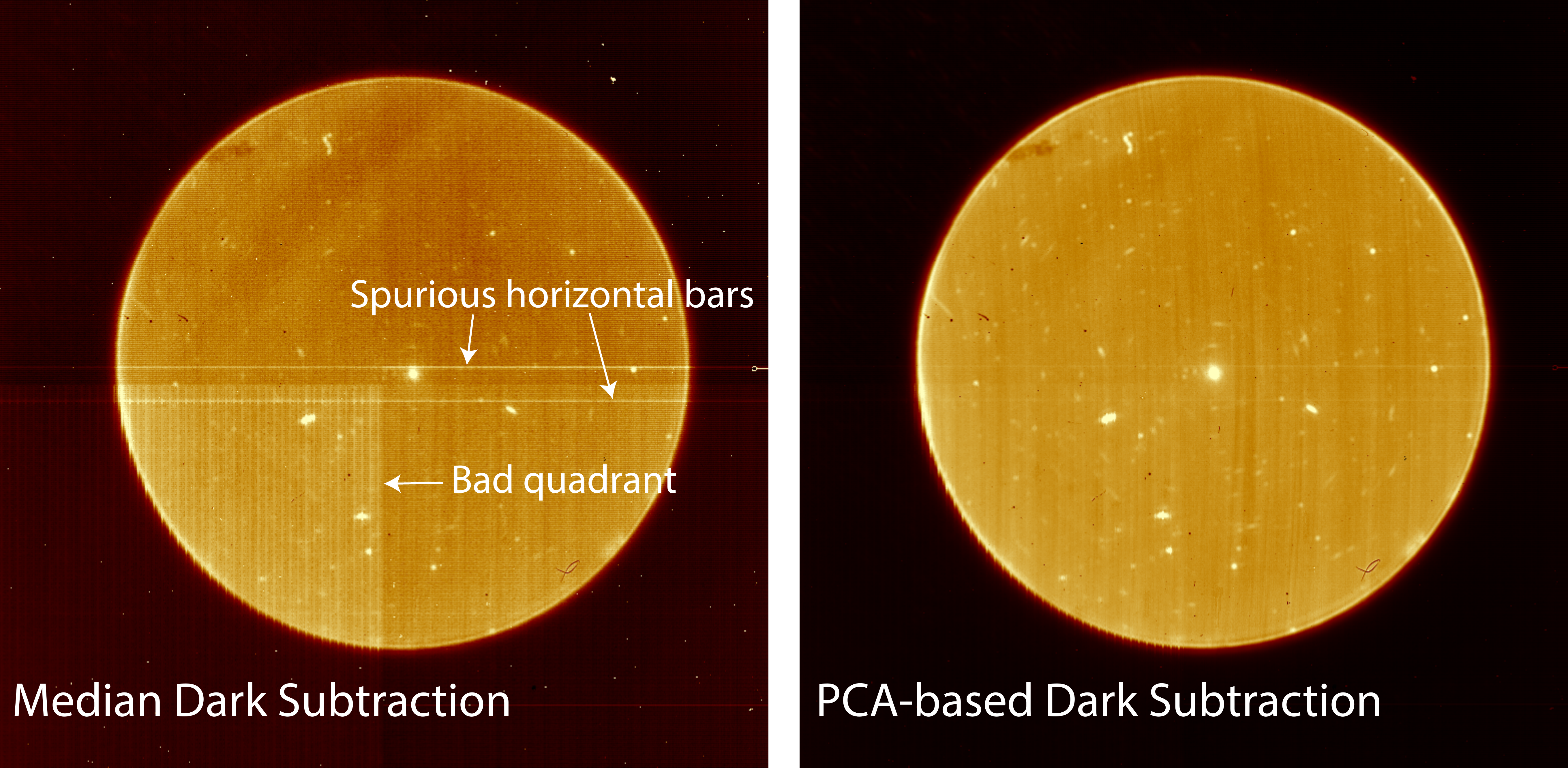}
	\caption{Median and PCA-based dark subtraction of a NaCo science cube captured in the middle of the CQ~Tau observation. Both are produced using the NaCo pipeline and are displayed on a linear scale with an intensity cut spanning 0.1--99.9 percentiles. The systematically bad quadrant has already been corrected as described in Section \ref{sec:naco}.} 
	\label{fig:pca_ds}
\end{figure}

We adapted the PCA-based sky subtraction routine in \textsc{vip} \citep[\texttt{cube\_subtract\_sky\_pca,}][]{GomezGonzalez2017} to model and subtract the varying (both spatially and temporally) bias level, dark current and thermal radiation from the instrument captured in the dark frames. In contrast to PCA-based sky subtraction \citep{Hunziker:2018ux}, the aim of PCA in this context is to best correct the readout grid pattern and systematic erroneous horizontal bars on the Aladdin 2 detector (Figure \ref{fig:pca_ds}, left) common in the dark and science frames. All dark frames are first cropped to the size of the science frames and saved into a cube for use as the PCA library. The systematic bad quadrant, AGPM centre and its shadow cast on the detector were masked to prevent potentially erroneous pixels and the stellar PSF being included in the PCA library. 

We begin an iterative loop over the science cubes to best correct the systematic horizontal bars: \textit{i}) add the difference of the median science pixel value and the median dark pixel value to the science cube; \textit{ii}) provide this science cube to \textsc{vip}'s \texttt{cube\_subtract\_sky\_pca} routine with the median dark cube as the PCA library to perform PCA dark subtraction with one principal component; \textit{iii}) restore the science image by adding the difference calculated in \textit{i}) to the PCA dark subtracted image; \textit{iv}) the standard deviation of pixel values in a subframe around the spurious horizontal bars is calculated; \textit{v}) loop over steps \textit{i--iv}) and search for the optimal offset (added in step \textit{i}) that minimises the standard deviation (measured in step \textit{iv}) using \texttt{scipy.minimize} (Nelder-Mead solver) that minimises the standard deviation metric; \textit{vi}) end the minimisation after 100 iterations or when the function converges on a minimum. We repeat this process for the sky and flat cubes.

Figure \ref{fig:pca_ds} presents a significant reduction of background bias in the corrected science frames when using PCA-based dark subtraction compared to a median subtraction. 

\section{Background Source} \label{sec:astrometry}
\begin{figure*}
	\centering
	\includegraphics[width=\linewidth]{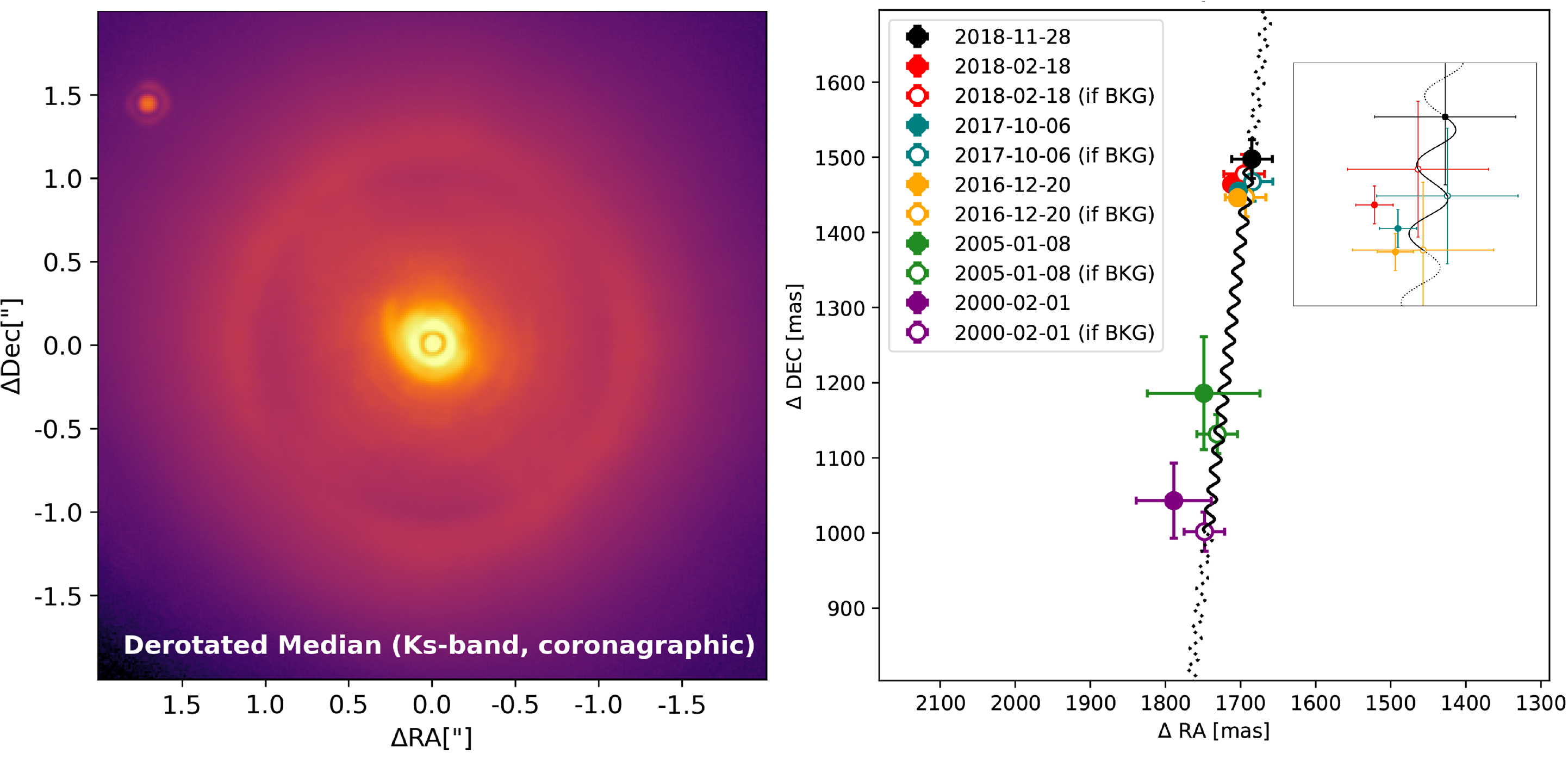} 
	\caption{\textit{Left}: Point-source detection in the 2016 IRDIS \textit{Ks}-band data set (north is up, east is to the left). Brightness is shown on a log scale spanning 0-99.9 percentile in intensity. \textit{Right}: Empty points correspond to the expected astrometry for a background object at each epoch, filled points are the measured astrometry. The astrometry is consistent with that of a background object.}
	\label{fig:background}
\end{figure*}
We resolved a point source at $\sim$2\rlap{.}\arcsec2 (327 au) separation from CQ~Tau in our ADI and PDI data sets. Figure \ref{fig:background} (right) shows the relative astrometry for the point source with respect to the star in the IRDIS ADI (Figure \ref{fig:background}, left), NaCo ADI and both IRDIS PDI data sets. The predicted track for a static background object is reversed through time, starting with the 2018 NaCo measurement.

In order to provide maximum time baseline, we considered archival NASA/ESO \textit{Hubble} Space Telescope (HST) observations from 2 February, 2000 (Program ID 8216) captured by the Wide Field and Planetary Camera 2 (PC2) instrument with the F675W ($\lambdaup$\textsubscript{$c$}~=~0.6717$\mu$m) and F814W ($\lambdaup$\textsubscript{$c$}~=~0.8012$\mu$m) filters. A second HST observation captured by the NICMOS instrument (camera 2) on 8 January 2005 with the F110W ($\lambdaup$\textsubscript{$c$}~=~1.025$\mu$m) filter was also considered. We conservatively assumed the uncertainty on each HST astrometric point to be the pixel scale of the corresponding instrument (0\rlap{.}\arcsec0455 pixel\textsuperscript{-1} for PC2, 0\rlap{.}\arcsec075 pixel\textsuperscript{-1} for NIC2) and assumed the star was at the intersection of the main diffraction artefacts in the PC2 dataset. The source was easily detected by eye in all data sets and its location was inferred by fitting a 2D Gaussian to a subframe encompassing the point source. In Figure \ref{fig:background} (right) we compare the detected motion of the source with the expected trajectory for a background star, considering the proper motion of CQ~Tau from \textit{Gaia} (RA = 2.98$\pm$0.06 mas yr$^{-1}$, Dec = --26.36$\pm$0.04 mas yr$^{-1}$, \citealt{Gaia-Collaboration:2021tu}). 

To explore whether this object previously interacted with the disc around CQ~Tau, we calculated its \textit{J--Ks} colour and inferred the spectral type using existing pre-main sequence \citep{Pecaut:2013va} and main sequence \citep{Ducati:2001} colours to be consistent with an M-type star. Knowing the contrast, apparent magnitude of CQ~Tau in \textit{Ks}-band (Section \ref{sec:detection_limits}) and the expected absolute magnitude for a young M-type star, we infer its distance to be $\gtrsim$ 2000 pc (compared to 149 pc for CQ~Tau). As the astrometry of the point-source is consistent with a background star with a small non-zero proper motion itself, we therefore rule out the possibility this is a bound object or stellar flyby.  

\appendix 



\bsp	
\label{lastpage}
\end{document}